\documentclass[12pt,preprint]{aastex}

\shorttitle{How to reach the orbital configuration?}
\shortauthors{Chen Yuan-Yuan et al.}

\begin{document}


\title{How to reach the orbital configuration of the inner three planets in HD 40307 Planet System ?}

\author{Chen Yuan-Yuan\altaffilmark{1}\altaffilmark{2}\altaffilmark{3}}
\author{Zhou Ji-Lin\altaffilmark{1}}
\author{Ma Yue-Hua\altaffilmark{2}\altaffilmark{3}}
\altaffiltext{1}{Department of Astronomy \& Key Laboratory of Modern Astronomy and Astrophysics in Ministry of Education, Nanjing University, Nanjing 210093, China; \url{zhoujl@nju.edu.cn}}
\altaffiltext{2}{Purple Mountain Observatory, Chinese Academy of Sciences, Nanjing 210008, China}
\altaffiltext{3}{Key Laboratory of Planetary Sciences, Chinese Academy of Sciences, Nanjing 210008, China}







\begin{abstract}

The formation of the present configuration of three hot
super-Earths in the planet system HD 40307 is a challenge to dynamical astronomers.
With the two successive period ratios both near and slightly larger
than 2, the system may have evolved from pairwise 2:1 mean
motion resonances (MMRs). In this paper, we investigate the
evolutions of the period ratios of the three planets after the primordial gas disk was depleted. Three routines are found to probably result in the
current configuration under tidal dissipation with the center star,
they are: (i) through apsidal alignment only; (ii) out of pairwise
2:1 MMRs, then through apsidal alignment; (iii) out of the 4:2:1
Laplace Resonance (LR) , then through apsidal alignment. All the three
scenarios require the initial eccentricities of planets $\sim0.15$,
which implies a planetary scattering history during and after the
gas disk was depleted. All the three routines will go through the
apsidal alignment phase, and enter a state with
near-zero eccentricities finally. We also find some special
characteristics for each routine. If the system went through
pairwise 2:1 MMRs at the beginning, the MMR of the outer two planets
would be broken first to reach the current state. As for
routine (iii), the planets would be out of the Laplace Resonance
at the place where some high-order resonances are located. At the
high-order resonances 17:8 or 32:15 of the planets c and d, the system will possibly enter the current state as the final equilibrium.

\end{abstract}





\section{Introduction}

Searching for Earth-size planets is one of the most exciting
objectives for the present exoplanet hunting. According to the
estimates from the results of the Kepler mission, more than 30\% of
the stars will host planets with mass less than 10 Earth masses
(so called super Earths), and the percent is even larger for the
planets around small-mass stars like M dwarf \citep{Howard2012,
Dressing2013}. Based on the core accretion planet formation theory,
super Earths were originally formed in distant orbits,
and migrate inward under the
interaction with the gas disk \citep{Ward1997, Tanaka2002,
Zhou2005}. Researchers found through numerical simulations that planets
are very likely to enter mean motion resonances (MMRs) during
the convergent migration processes \citep{Terquem2007}.
Some exo-
planet pairs are observed to be in exact MMRs,e.g., GJ 876b,c,d \citep{Rivera2010}.
However, the statistics from Kepler planet candidates shows
that most of the planet pairs are in near MMRs with period ratios slightly
larger than the exact integer ratios \citep{Fabrycky2012, Batygin2013}.

Several mechanisms have been proposed to explain the origin of those
period ratios slightly larger than the integer ratios. For short
period planets, star-planet tidal interactions would deplete the
semi-major axis of the inner planet more quickly and make the period
ratio dispersing \citep{Lithwick2012, Batygin2013}. For long period
planets, \citet{Papa2013} proposed that the gravitational interactions
between partial gap-opening planets and the gas disk may instead
provide efficient dissipation. Besides, \citet{Lee2013} raised that tides between planets and star are not strong enough to increase part of the planet pairs to the current separations, and it is also uncertain that the migration of planets in the disk could result in the observed distribution of planetary period ratios because of the complex disk environment. \citet{Wang2012} worked on the formation of a near Laplacian resonance configuration in the KOI-152 system, and focused on the influence of stellar accretion, stellar magnetic field and the speed of migration in the protoplanetary disk.

HD 40307 system plays a notable role and its migration history
has been studied in detail representatively. The three inner planets
in this system have the period ratios 2.23 and 2.13 for the inner and outer
pairs, respectively \citep{Mayor2009}. Both \citet{Zhou2010} and \citet{Papa2010}
(PT2010 hereafter) simulated their inward migration in a gas disk,
and found that the three planets were trapped into pairwise 2:1 MMR
or Laplace Resonance (LR) or both of them, and retained in it until the tidal dissipation between star and planets is effective, which may drive  it out of the resonances. However, how the exact final configuration is achieved under different initial conditions (pairwise 2:1 MMRs or LR, etc.) is not fully understood. PT2010 also demonstrated that in low-eccentricity situation, the period ratios evolve too slowly to reach the current orbital
architecture. Meanwhile, they claimed that the current state of the
system cannot be originated from LR directly.

\begin{figure}
\centering
\includegraphics[width=14cm]{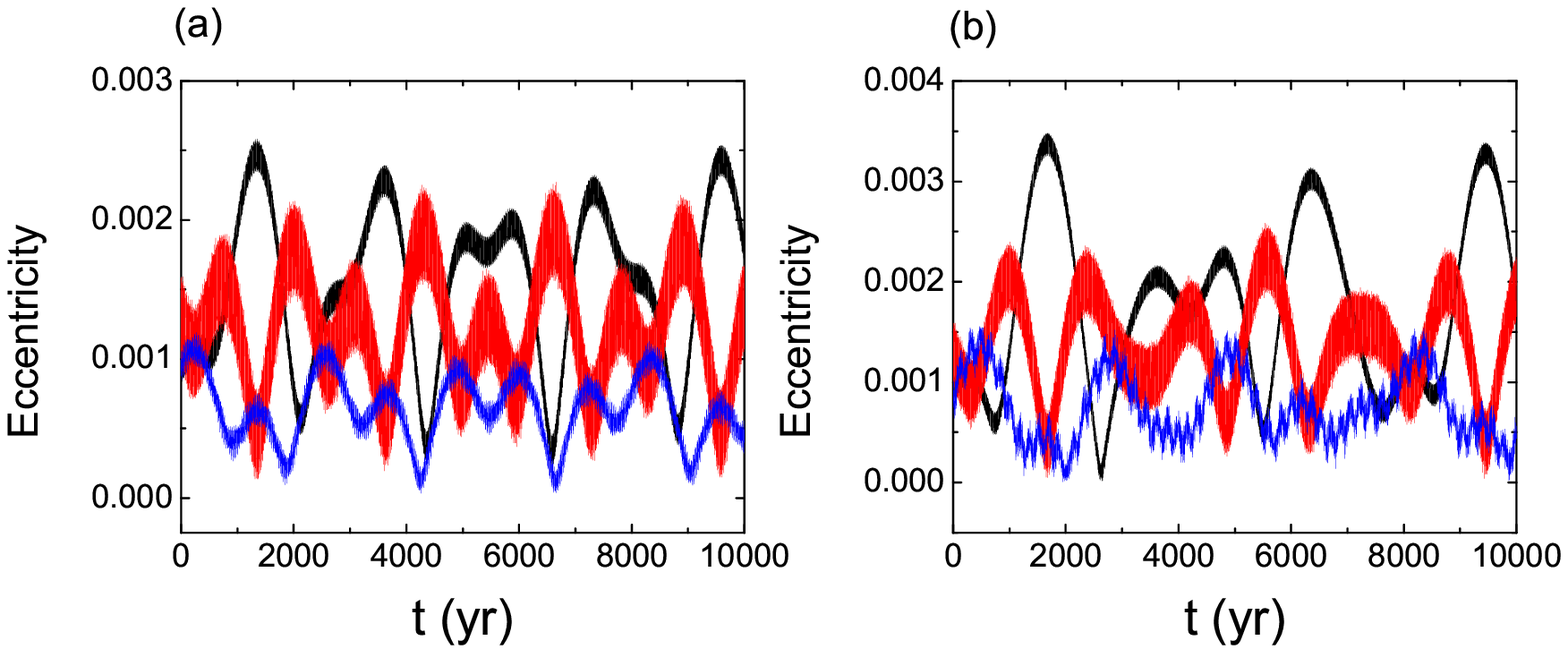}
\caption{The evolution of the eccentricities of the planets
HD40307~b,c,d, without (a) and with (b) the outer
planets HD40307~e,f,g.  \label{6planet}}
\end{figure}

Recently, \citet{Tuomi2012} reanalyzed the RV data of the system HD40307, and claimed that it exits three additional planets (e,f,g) with masses of $3.5m_E, 5.2m_E$ and $7.1m_E$ at orbits with periods of $34.62d, 51.76d$ and $197.8d$, respectively. If the presence of additional planets is confirmed, the planet system is very compact especially for the inner 5 ones.
To reveal whether the additional outer planet might interact with the inner ones, we calculate the relative space among planets, and find that there are respectively 21, 17, 13, 11, 33 times of their mutual Hill's radii among the 5 neighboring pairs. So dynamically the planets d,e,f are more closely related. Assuming they have around 10 Earth masses, the orbital crossing time could be around $10^{8-9}$ times of their periods, i.e., $10^{7-8}$ years (See Eq. (3) or Figure 3 of \citet{Zhou2007}). This time scale is comparable to the above tidal evolution for $Q'=100$. So if the outer three planets are confirmed, the inner five planets could evolve as a dynamically related system. However, since our discussed configurations are from MMRs, and only the inner three are effected significantly by tidal dissipations, the above three routines might not change too much unless the outer planets are also involved in MMRs. Meanwhile, Figure ~\ref{6planet} gives the comparison of the eccentricities of the planets HD 40307 ~b,~c,~d. We can see the secular oscillations of the eccentricities are in the same order of magnitude whether the outer three planets are considered or not. According to these, the model we apply includes the inner three planets merely, which reduces the number of degree of freedom and simplifies the problem dramatically.

In this paper, we reconsider the process of tidal dissipation and study the final configurations of the inner three planets under different evolution scenarios. Three types of evolution routines have been explored to get to the observed state finally. We deduce that the apsidal alignment phase is a very common and maybe a necessary phase for such compact systems under tidal dissipation.
In Section 2, we describe the numerical model. Section 3 discusses the
possible current configuration of the three planets. Section 4
describes the three kinds of evolution paths which can get to the
current configuration. Then we compare the three evolution paths and make
some speculations in Section 5. The last section gives a summary of this paper.

\section{Modal}

We consider a host star and N planets in a barycentric inertial
coordinate system (N=3 for HD 40307 system). There are three
additional forces besides gravitational interactions. The equations
of motion are:
\begin{equation}
\frac{\emph{d}\,^2\textbf{r}_i}{\emph{dt}\,^2}=\sum_{j=1,j\ne
i}^n\frac{Gm_j(\textbf{r}_j-\textbf{r}_i)}{\mid\textbf{r}_j-\textbf{r}_i\mid^3}+\textbf{\textit{f}}_{TD}+\textbf{\textit{f}}_{QD}+\textbf{\textit{f}}_{GR}\qquad
i=0,\dots,3,
\end{equation}
where i=0 represents the star, and i=1,2,3 represnets the
planets b,c,d respectively.
$\textbf{\textit{f}}_{TD},\textbf{\textit{f}}_{QD}$ and $\textbf{\textit{f}}_{GR}$
denote the acceleration produced by tidal damping, quadrupole moment
and general relativistic respectively (\citet{Mardling2002}). The specific expressions are
\begin{eqnarray}
\textbf{\textit{f}}_{TD}&=&-\frac{9n}{2Q'_p}(\frac{m_0}{m_p})(\frac{S_p}{a})^5(\frac{a}{r})^8 \nonumber \\
&&\cdot[3(\hat{\textbf{r}}\cdot\dot{\textbf{r}})\hat{\textbf{r}}
+(\hat{\textbf{r}}\times\dot{\textbf{r}}-r\mathbf{\Omega}_p)\times\hat{\textbf{r}}]\\
\textbf{\textit{f}}_{QD}&=&\frac{S_p^5(1+m_0/m_p)k_p}{r^4}\left\{{\left[5(\mathbf{\Omega}_p\cdot\hat{\textbf{r}})^2-\Omega^2_p
-\frac{12Gm_0}{r^3}\right]}\right. \nonumber \\
&&\left.{\cdot\hat{\textbf{r}}-2(\mathbf{\Omega}_p\cdot\hat{\textbf{r}})\mathbf{\Omega}_p}\right\}\\
\textbf{\textit{f}}_{GR}&=&-\frac{Gm_{0p}}{r^2c^2}\left\{{\left[{(1+3\eta)\dot{\textbf{r}}\cdot\dot{\textbf{r}}
-2(2+\eta)\frac{Gm_{0p}}{r}}\right.}\right. \nonumber \\
&&\left.{\left.{-\frac{3}{2}\eta\dot{r}^2}\right]\hat{\textbf{r}}-2(2-\eta)\dot{r}\dot{\textbf{r}}}\right\}
\end{eqnarray}
where $m_p, S_p, \mathbf{\Omega}_p$ is the mass, radius, and spin speed of planet, respectively.
$m_{0p}=m_0+m_p,\;\eta=m_0m_p/m_{0p}^2$. $\textbf{r},\dot{\textbf{r}}$ is the position vector and speed vector of planet relative to the central star. $c$ is the speed of light. $Q'_p=3Q_p/(2k_p)$.

As the tidal dissipation
from the planets deformation is much bigger than that from the
star's, we consider the planetary tide merely. The minimum masses of all planets in this system have the same order
of magnitude with the Earth, so we suppose the same damping
parameter $Q_p=0.01$ and apsidal motion constant $k_p=0.3$ \citep{Zhou2008,
love2009}for all three planets in all simulations. Terrestrial planets have $Q_p=10-100$ \citep{Goldreich1966} and we set a smaller value to accelerate the damping and shorten the calculating time as \citep{Mardling2007} and PT2010 did. Besides, we get
the radii of planets by supposing the densities of the planets equal
to Earth's. In fact, the radii are coupled with the tidal
dissipation parameter $Q'_p$ in the expressions of all additional
forces, and $Q'_p$ is inversely proportional to tidal damping
timescale, so some deviations of radii or $Q'_p$ are equivalent to a
change of evolution time in most cases.

The timescale of planetary rotation during tidal damping is much
shorter than that of orbital evolution, so we set
$\Omega_p=n$ and the spin axis is perpendicular to the orbital
plane at the beginning. Subsequent evolution of $\mathbf{\Omega}_p$
is given by the relation as follow \citep{Mardling2002}
\begin{equation}
\dot{\mathbf{\Omega}}_p=-\frac{m_0m_p}{I_p(m_0+m_p)}\textbf{r}
\times(\textbf{\textit{f}}_{TD}+\textbf{\textit{f}}_{QD}),
\end{equation}
which is deduced by the conservation of total angular momentum, and
$I_p$ is inertial moment of planet.

We apply the RKF78 variable-step integrator to make the N-body simulations, and the additional forces are added during every step. The numerical error for every step is set to be $10^{-12}$, and the total energy is generally conserved to $10^{-6}$ in the conservative cases \citep{Ji2005}. The integration time is about 10-12 hours for every run. The elements of planets are output in equal interval (every 100yrs) to track the evolutions.

\section{The state of the three planets HD40307 ~b,~c,~d in low-eccentricity situation}

We first set small eccentricities, and make the simulations with the
observed semi-major axes and minimum masses (Table \ref{initial})
(We adopt the minimum masses of the planets as fiducial values in our
simulations, and the effects of more massive planets will be
discussed in section \ref{discuss}), and altering initial
eccentricities and phase angles. The resonance angles are
given by
$\Phi_1=2\lambda_2-\lambda_1-\varpi_1,\Phi_2=2\lambda_2-\lambda_1-\varpi_2,
\Phi_3=2\lambda_3-\lambda_2-\varpi_2,\Phi_4=2\lambda_3-\lambda_2-\varpi_3$ \citep{Lee2002, Ji2002}.
Here $\lambda_i$ and $\varpi_i$ represent the mean longitude and the
longitude of pericentre of planet i, and the indices $i=1,2,3$ stand
for the planet b,c,d, respectively.

\begin{table}[H]
\label{initial}
\caption{Orbital elements of HD40307 b,c,d and the star from
\citet{Mayor2009}.}
\begin{center} \footnotesize
\begin{tabular}{cccccc}
\hline\hline
Parameter &  & HD40307~b & HD40307~c & HD40307~d \\
\hline
$m_2\sin{i}$ & [$M_{\oplus}$] & 4.2 & 6.9 & 9.2 \\
$P$ & [days] & $4.3115\pm0.0006$ & $9.620\pm0.002$ & $20.46\pm0.01$ \\
$a$ & [au]   & 0.047    &  0.081    &  0.134 \\
$e$ &      &   0.0     &  0.0       & 0.0 \\
\hline\hline \
Star & Mass & Sp. type & Metallicity & $T_{eff}$ \\
 HD40307 & [$M_{\odot}$]  &  & [dex]  &  [K] \\
\hline
 & $0.77\pm0.05$ & K2.5V & $-0.31\pm0.03$  & $4977\pm59$  \\
\hline\hline
\end{tabular}
\end{center}
\end{table}
\vspace{0.0mm} {\footnotesize}\vspace{0.0mm}

\begin{figure}[H]
\centering
\includegraphics[scale=0.25]{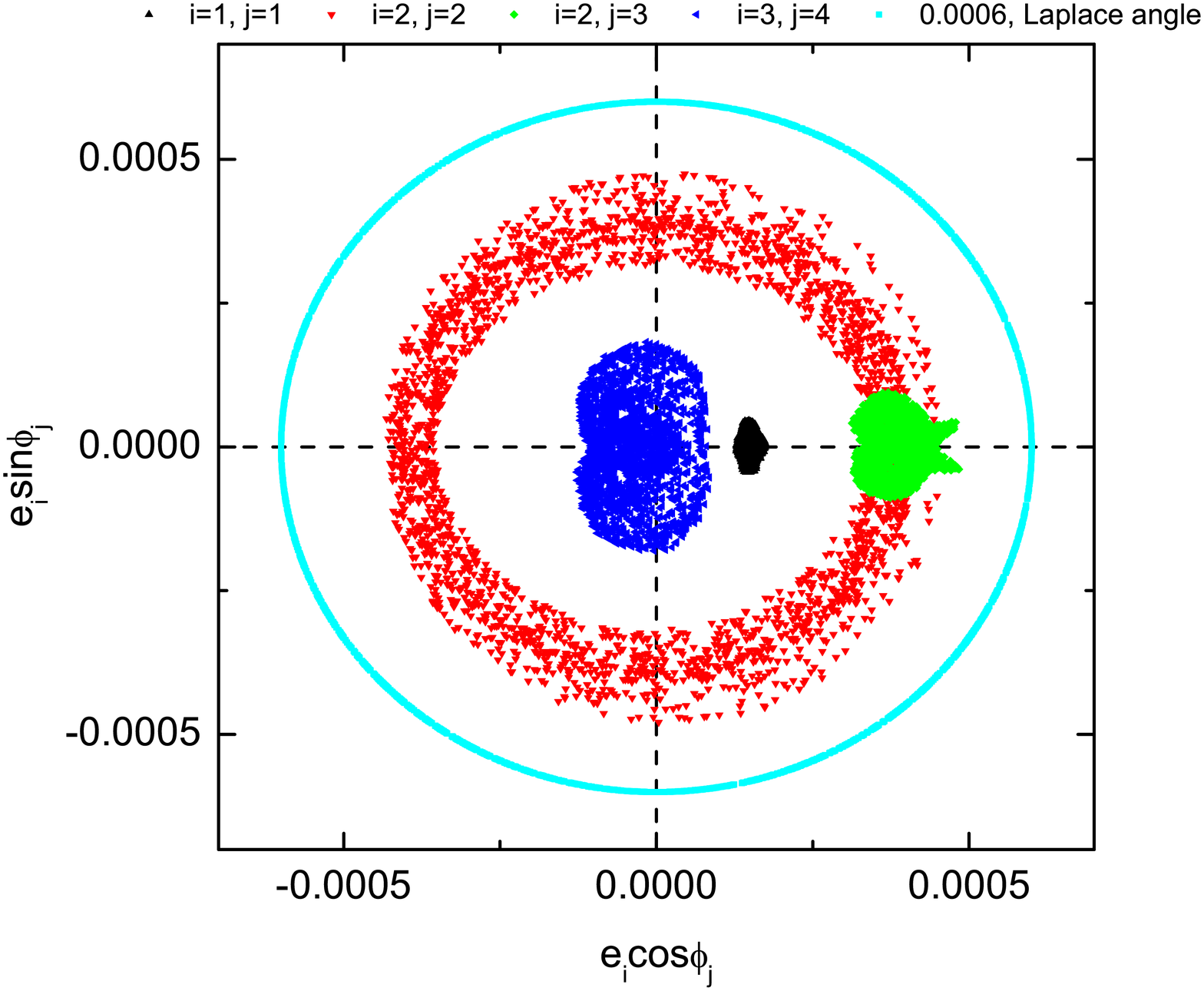}
\caption{The equilibrium state of
HD40307 system under tidal dissipation. It shows that $\Phi_1$(black
dots),$\Phi_3$(green dots) librate around 0, and $\Phi_4$(blue dots)
nearly librates around $\pi$. The eccentricities in the equilibrium
are in the order of magnitude of $10^{-4}$. Laplace angle(cyan dots) is
circulating. \label{current}}
\end{figure}

Then we find that as long as the eccentricities are small ($\sim
10^{-4}$), whatever initial phase angles are set to be, the system
would eventually come to the same equilibrium state, with
$\Phi_1$,$\Phi_3$ librating around 0, and $\Phi_4$ ``nearly
librating'' around $\pi$ (``nearly librating'' here means that the
resonant angle has no obvious libration but is just more dense
around some place and has long-term time average, or is associated
with long term changes of the orbital elements). Figure
~\ref{current} shows the equilibrium state in the $e-\Phi$ phase
space. And it is consistent with the simulations in PT2010.

\citet{Delisle2012} gave a global analysis of the phase space of the
situation above, and demonstrated that the apparent libration of the
resonant angles in low eccentricities situation results from the
severe damping of the amplitudes of the eigenmodes in the secular
motion, and the planets are not really in the MMRs. Indeed, under tidal dissipation, the separatrices that exist in the resonant systems eventually disappear when the eccentricities of planets are very small. There is only a circulation of the orbits around a single elliptical fixed point left in the phase space (see Fig. 2 in \citet{Delisle2012}).

\section{Paths that will evolve to the present configurations}

After the gas disk disappears,
tidal dissipation between the star and planets will dominate the migration of planets in close-in orbits, which basically cause inward migration
when the planets are inside the synchronous orbit of stellar spin.
The decayed timescale of the semi-major axis due to planetary tides can be estimated as
$\tau_a=\tau_{\rm circ}/ e^2$ \citep{Lithwick2012, Zhou2008}, where $\tau_{\rm circ}$ is the orbital circularization time scale due to tide, and $e$ is the orbital eccentricity of the planet. So different eccentricities and the
relative magnitudes of three planets' eccentricities would
correspond to different evolution directions of the period ratios of
the adjacent planets. And the orbital eccentricities are mainly
determined by the orbital configurations and the resonance types.

In this section we focus on the moderate-eccentricity situation and
investigate different configurations and resonances that the planets may have
gone through. As collisions or scattering are very likely to take
place during the gas disk dissipation for such a compact system, high
eccentricities would be common before tidal damping effects \citep{Ogihara2010}. PT2010
pointed out that if the eccentricity of the outermost planet is
up to 0.15, the period ratios would shift to the nearby current
values from the values close to 2. Our simulations are generally
consistent with PT2010. Further more, we make comparisons
and classifications, and find three kinds of paths which can get to
the current state from different initial states.

{\em Path 1: Apsidal alignment}.

\begin{figure}[H]
\centering
\includegraphics[scale=0.45]{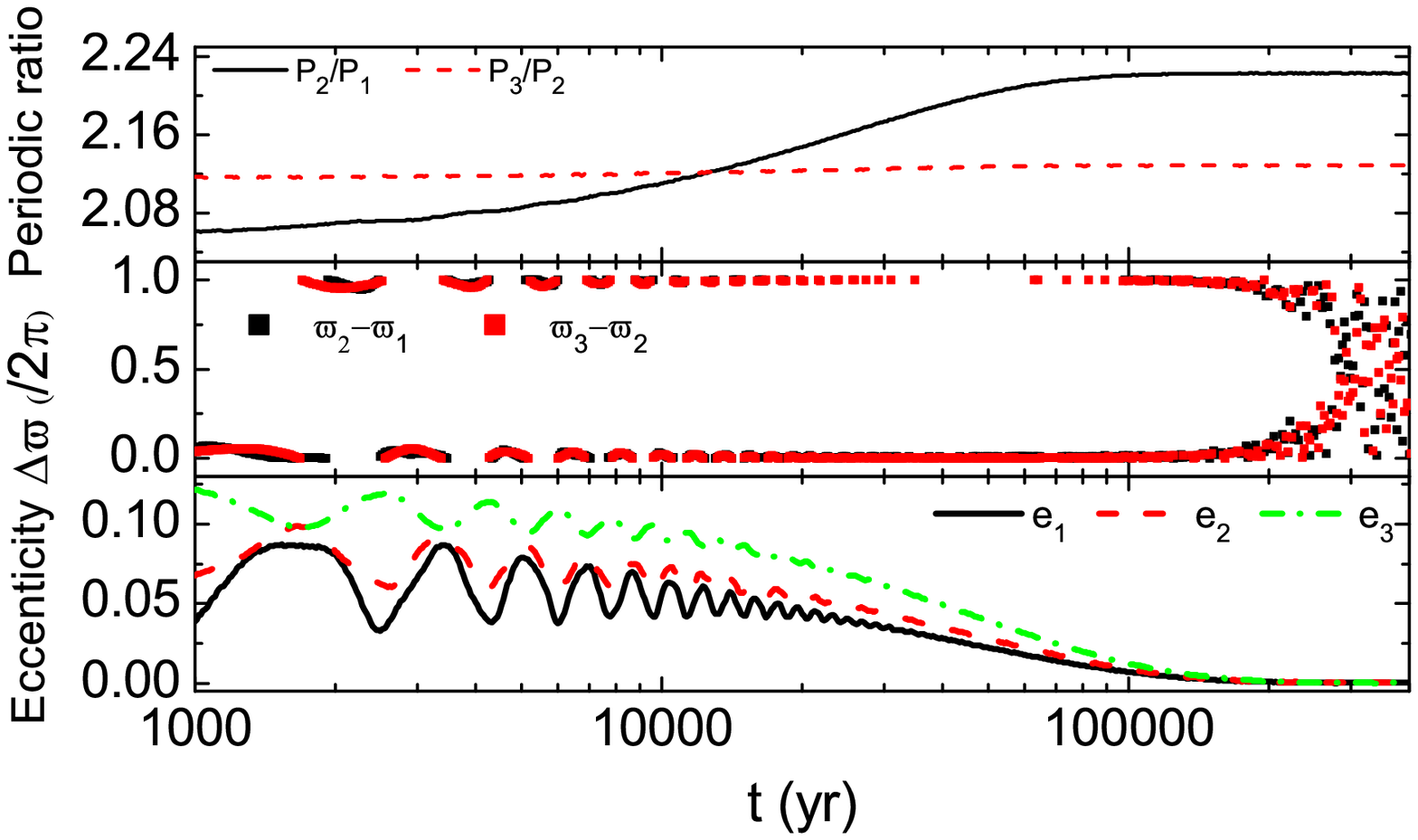}
\caption{Period ratio (top),
differences of longitude of pericentre (middle) and eccentricities
(bottom) evolve with time in one simulated run. Three planets are
initially located at 0.05 au,0.08 au and 0.134 au ($P_2/P_1=2.055,
P_3/P_2=2.12$), with the same eccentricities 0.1. The phase angles are
chosen arbitrarily. The apsidal alignments($\Delta\varpi\approx 0$)
are kept until the eccentricities are damped to very small values.
\label{secular1}}
\end{figure}

Assume that the three planets were formed around the moderate region
($\gtrsim 1$ AU) one after another, and the first-formed planet
migrated inward first, and then the three planets may have a history
that were far away from any MMRs. After the gas disk disappears,
they would undergo secular evolution under their mutual interactions.
To investigate this type of evolution with emphases on the final
configuration, we fix the initial conditions $P_2/P_1=2.055$,
$P_3/P_2=2.12$, $e_1=e_2=e_3=0.1$, so that the orbits can evolve to
around the present configuration. The evolution is shown in
Figure~\ref{secular1}.

\begin{figure}[H]
\centering
\includegraphics[scale=1]{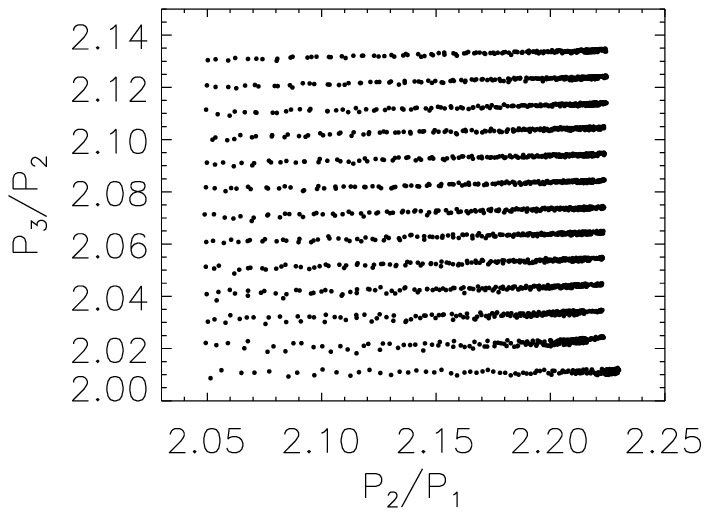}
\caption{Evolution tracks of
13 orbits with $P_2/P_1=2.01,2.02,...2.13$ are plotted in the
$P_2/P_1-P_3/P_3 $ plane. Other initial conditions are $P_3=20.5$
days, $P_3/P_2=2.05$, $e_1=0.15$, $e_2=e_3=0.0001$ . Angular parameters
are set randomly. \label{secular2}}
\end{figure}

One of the major characteristics for this evolution under tidal
dissipation is that the planets will be quickly driven into three-body
secular phase locking, or apsidal alignment ($\varpi_1 \approx
\varpi_2 \approx \varpi_3 $). The reason is that, orbital alignment
(i.e.,  $\Delta\varpi =0$) is a quasi-equilibrium state in the
$e-\Delta\varpi$ plane (see Fig. 1 of \citet{Mardling2007}, also \citet{Zhou2003} for the non-dissipation cases). Though
the planets in our cases are too close to be approximated by a
hierarchical system, the evolution scenario here is quite similar
to that in a hierarchical system, as shown in \citet{Mardling2007}. First, the three planets align quickly. During
the alignment process, the amplitudes of the oscillations of
eccentricities decrease to $\sim 10^{-4}$ within $3\times 10^5$
years for $Q_1=0.01$, which corresponds to $~3$ Gyrs provided
$Q'=100$. The alignment configuration is kept until the eccentricities are damped
to almost 0. Then the alignments of apsidal lines are broken, and
the system turns to the low-eccentricity state (the same as shown in
Figure ~\ref{current}).

Another feature of the evolution path is that $P_3/P_2$, the period
ratio of the outer pair, has no significant change in this process. Thus
the evolution track in the $(P_2/P_1, P_3/P_2)$ plane is almost a
line parallel to the x-axis until the end of evolution. Figure
~\ref{secular2} emphasizes the feature further. It shows the evolution
tracks of 13 orbits with different initial $P_1$ according to $
P_2/P_1=2.01,2.02, 2.03, ..., 2.13$. Why does $P_3/P_2$ change
slightly in these evolution paths? On the one hand, orbital angular
momentum and energy transfers among different planets are much
weaker than that in any two-body mean motion resonance or three-body
resonance. On the other hand, the tidal dissipation of the middle
and outermost planets is not obvious as the planets are not close enough
to the star. Accordingly, we infer that the period ratio $P_3/P_2$
should have approached the current value before the system began the apsidal-alignment evolution. So this path could just be as
an intermediate stage if the system was in 2:1 MMRs before tidal
evolution.

\begin{figure}[H]
\centering
\includegraphics[scale=0.45]{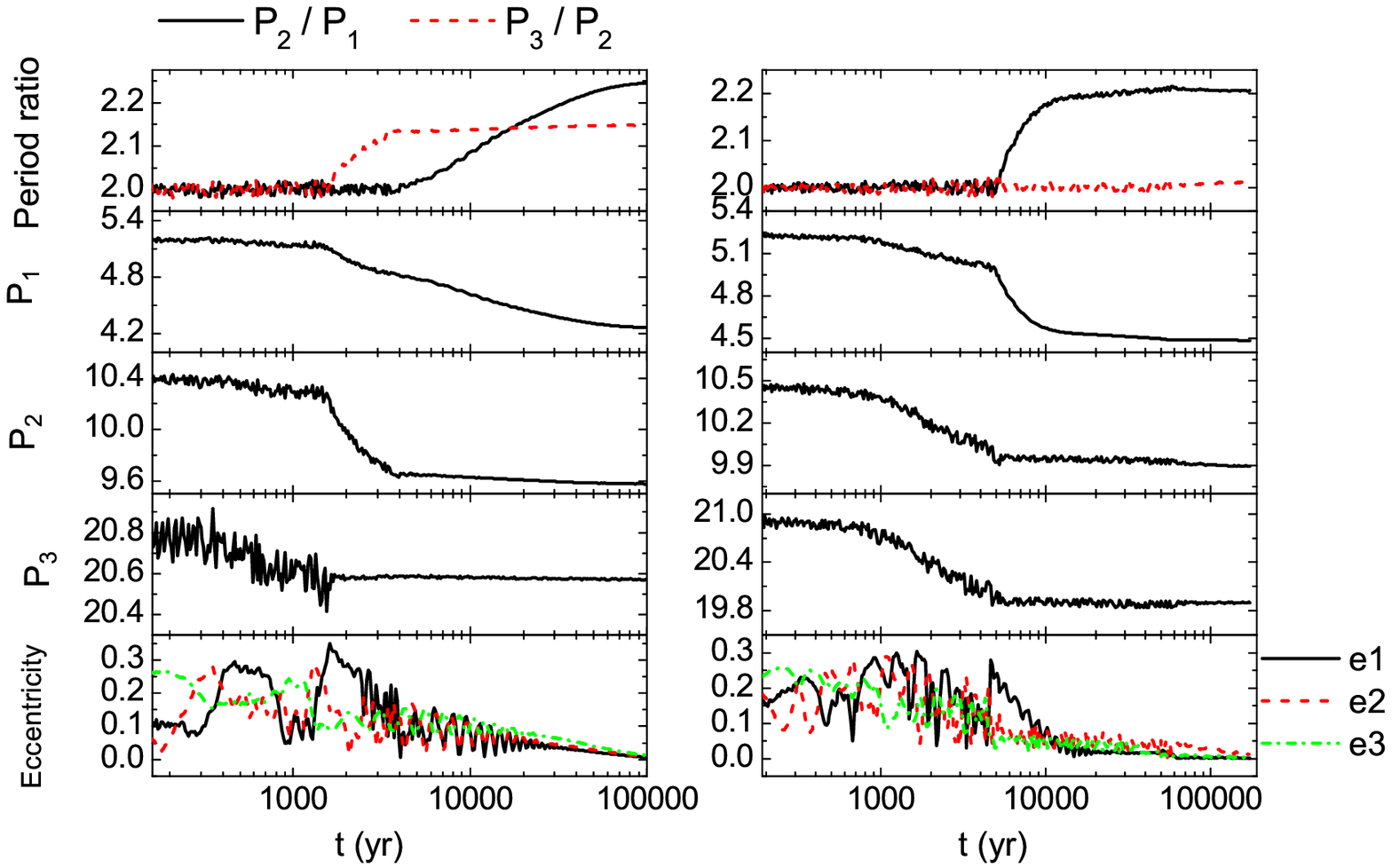}
\caption{Two examples for three planets initially in the pairwise
2:1 mean motion resonances with high eccentricities. Each column
shows one example. In the left, three planets are put at 0.054,0.085
and 0.136 au respectively ($P_2/P_1=P_3/P_2=2.0001$), with
$e_1=0.07,e_2=0.25,e_3=0.19$. In the right, only the inner planet's
initial eccentricity is different, $e_1=0.19$. In the left case, the
outer pair of planets are first out of 2:1 MMR , which causes the two
period ratios reaching up to around the present position finally.
\label{mmr21}}
\end{figure}

{\em Path 2: pairwise 2:1 MMRs.}

Migration of planets embedded in the protoplanetary disk is very
common \citep{Goldreich1979, Ward1986, Ward1997, Tanaka2002}.
\citet{Zhou2010} and PT2010 specifically simulated the three planets
in HD40307 system migrating in the gas disk, and both found that the
planets are easily trapped into pairwise 2:1 MMRs or Laplace
Resonance during the migration. Assuming the initial configurations
$P_2/P_1\approx 2, P_3/P_2\approx 2$, we investigate the subsequent
evolution of three planets under tidal effects with the star.

Figure \ref{mmr21} shows two typical orbits with different initial
eccentricity $e_1$, and one of them reaches the observed state
finally. The resonances are very unstable and are disrupted
$10^3-10^4$ years later with Q'=0.01 (corresponds to $10^7-10^8$
years for $Q=100$). The two examples show mainly two types of
breakup of MMRs: in the left case, the outer 2:1 MMR goes out first,
and then the middle planet continues to be dragged in by the innermost
planet because of the inner 2:1 MMR, which makes $P_3/P_2$
increasing quickly to the present value. Instead, if the inner pair
of planets go out of the 2:1 MMR at first, like the case in the
right, then $P_3/P_2$ would keep around 2, because without the inner
2:1 MMR, the middle planet could not move inward more than the
outermost planet. So for this kind of path, the outer pair being out
of MMR first is the necessary precondition for the system coming to
the present configuration.

{\em Path 3:  Laplace Resonances.}

Two successive period ratios both approaching 2 also remind
us whether the three planets are in LRs.
The LR is defined as $n_1-3n_2+2n_3\approx0$ so that
$\Phi_L=\lambda_1-3\lambda_2+2\lambda_3$ liberates around either
$0^o$ or $180^o$. Such a configuration has been discovered and
investigated among the Galilean satellites of
Jupiter \citep{Peale2002}. The satellites go through either the
primordial inward migration due to interactions with a circumjovian
disk, or subsequent differential orbital expansions from tides
raised on Jupiter, and then were trapped into pairwise 2:1 MMRs as well
as LR with $\Phi_L$ liberating around $\pi$.

\begin{figure}[H]
\centering
\includegraphics[scale=0.5]{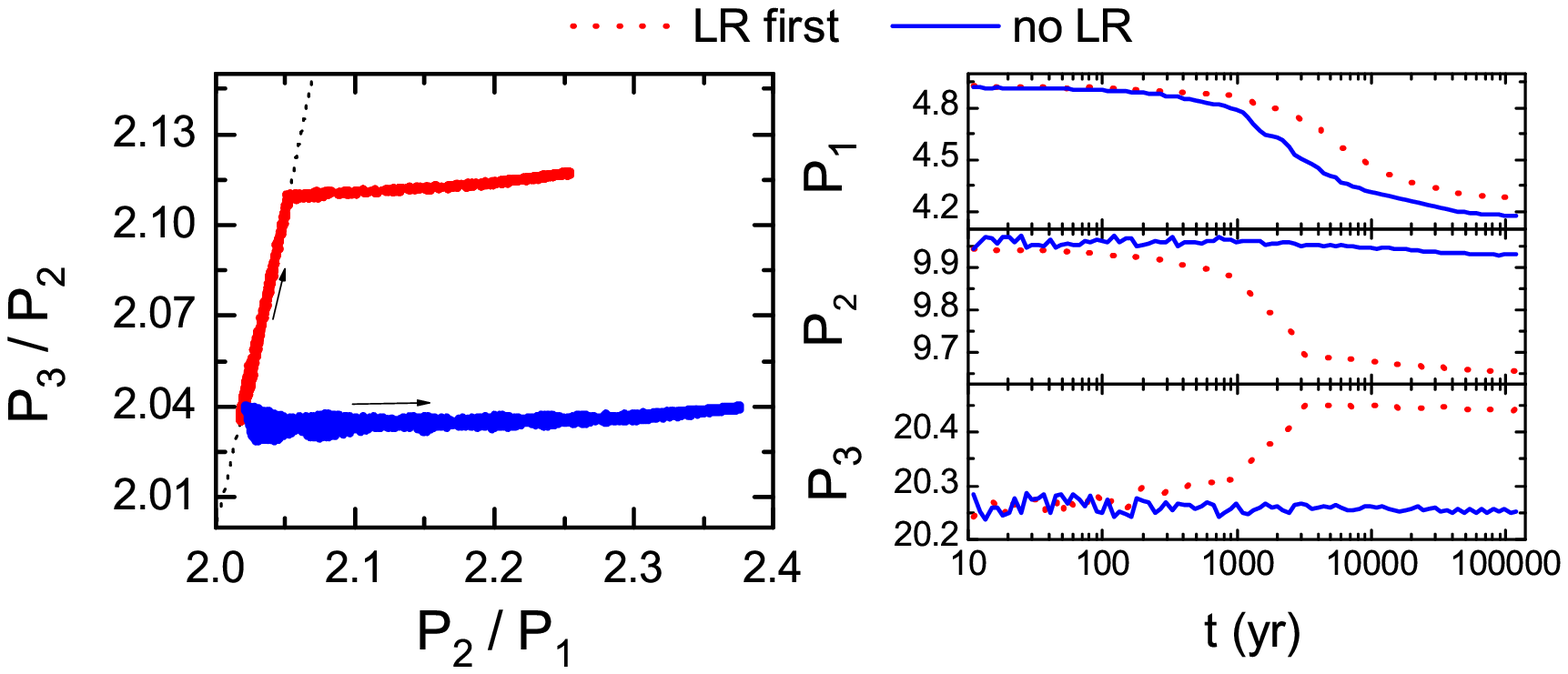} \caption{The comparison of two
cases originally in the Laplace relation. The two cases have the
same parameter, $P_2/P_1=2.02, P_3/P_2=2.04$,
$P_3=20.2865$days, $e_1=0.19$, $e_2=0.19$, $e_3=0.01$, with
different phase angles. The left panel is the period ratio tracks,
and the right ones are the period evolutions of three planets versus
time. \label{laplace}}
\end{figure}

PT2010 has clarified a negative conclusion based on the planets' mean
motions not satisfying the Laplace relation $3n_2-2n_3-n_1=0$. In
spite of that, we find LR still a possible part of
evolution process, and such a case is given in Figure \ref{laplace}.
Two cases in the figure have the same initial conditions except for
phase angles. The case in dash line enters LR at first, and the
period ratios go along the Laplace relation at this duration (the
dot black line). Then LR is broken at some place, and secular
evolution follows before a tidal equilibrium comes. Compared
to this, with the different initial phase angles, the case in solid
starts secular evolution directly, and the final equilibrium is
far away from the current. In the case, the periods of the outer two
planets hardly change during the evolution, while they vary a lot in
the LR case due to the strong interaction among planets. Hence, LR
trapping becomes the key step to shift $P_3/P_2$, and propel the
system into the current under this kind of initial conditions.

\begin{figure}[H]
\centering
\includegraphics[scale=0.5]{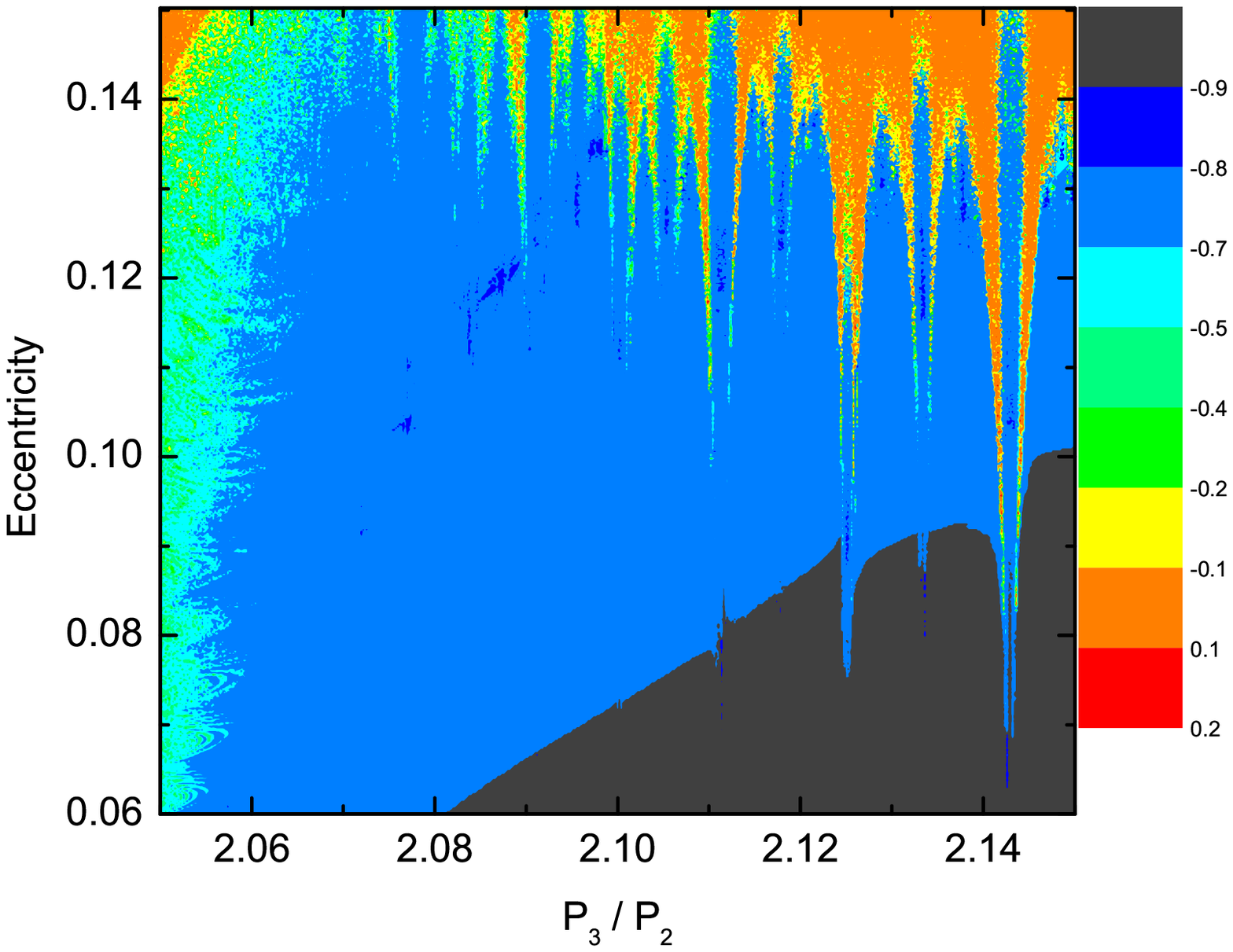}
\caption{Contour of the oscillating amplitudes of the Laplace angle.
Blue end of the color bar represents libration, and red end means
circulation. Every case is integrated $10^5$ yrs without tidal
dissipation. As for the initial conditions, the outermost planet is
fixed at 20.8 days, $P_2/P_1$ is calculated using $P_3/P_2$ and the
Laplace relation. Three planets have the same initial eccentricity
for reducing the variations. Initial phase angles are set as
$\varpi_1=0,\varpi_2=\pi,\varpi_3=0,\lambda_1=0,\lambda_2=\pi,\lambda_3=0$,
which can make the system enter LR easily. From the panel,
we can see some more unstable place of LR on $P_3/P_2\simeq17/8,
32/15, 15/7$. These places should be the reason why
Laplace Resonance is broken in the dash case of Figure \ref{laplace}.
\label{xn23_e}}
\end{figure}

Trapping into LR is a quite stochastic event, and mainly depends on
the phase angles of planets at the moment when the Laplace relation is
satisfied. However, there seems to be some rules to follow the place where the system is out of LR. We made a scan on the oscillating amplitude of the Laplace angle for different period ratios and different
eccentricities, by N-body simulation without dissipation (Figure
\ref{xn23_e}). We found that for the same eccentricities, there are
some places where LR is more unstable, such as $P_3/P_2\simeq2.125,
2.143,...$, which corresponds to the high-order resonances $17/8,
15/7,...$ of the outer two planets. Under tidal dissipation,
$P_3/P_2$ increases and eccentricities decrease, so the
corresponding position of the state in the $P_3/P_2-e$ phase space
will move toward the lower-right and encounter a series of the
high-order resonances. For the cases approaching the current
state finally, 17/8 or 32/15 would be the high-order resonance from
which the system comes out of the LR. Because after out of the LR,
the system will enter the apsidal-alignment state, in which the
outer period ratio $P_3/P_2$ will not change a lot.

\begin{figure}[H]
\centering
\includegraphics[scale=0.5]{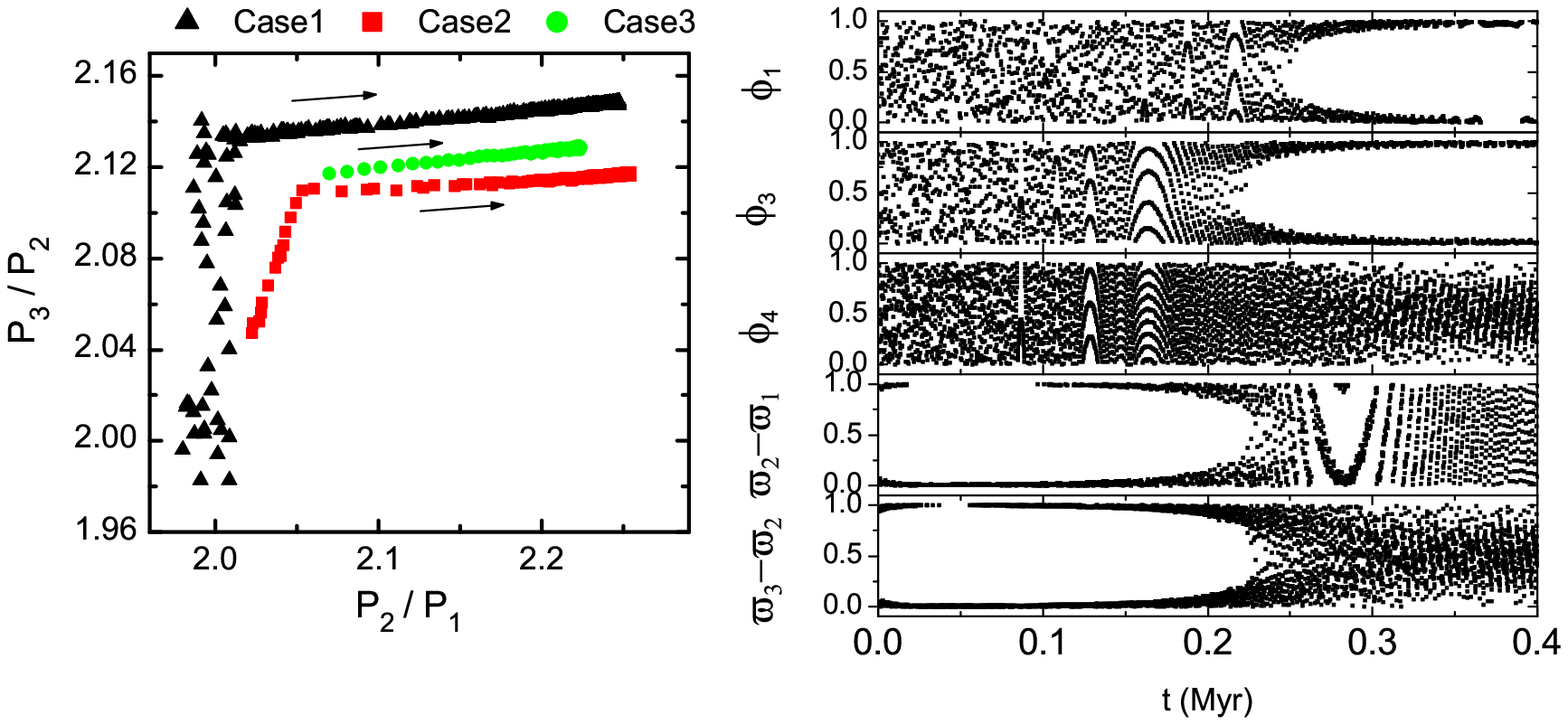} \caption{The left panel shows the
paths of $P_2/P_1$ versus $P_3/P_2$ of three representative cases,
which are originated from different initial conditions, and get to
around the current state of HD40307 finally. All of the paths have a
parallel part, which corresponds to a secular evolution with apsidal
alignment. Moreover, the runs turn into the same state at the end,
which is also the one in Figure ~\ref{current}. The right three
panels show the resonance angles and differences of longitude of
pericentre of Case 3 versus time. For the other two cases, these
phase angles evolve similarly when they enter the apsidal
alignment part, and are not shown here. \label{long}}
\end{figure}

\section{Comparison and some speculations}
\label{compair}

We compare the states the system goes through in the three kinds of
paths, and find that although the planets stem from different
states, they all include the processes of apsidal alignment and the
following low-eccentricity equilibrium (Figure ~\ref{long}). The
final equilibrium state has no difference in the three kinds of
paths, so we still cannot tell the exact story even though the
eccentricities or the resonance angles have been detected precisely.
However, due to the secular evolution as a common state in these
evolution histories (the horizontal part of paths in Figure
~\ref{long} left), if the current eccentricities are $\sim10^{-4}$
as figure ~\ref{current} shows, it would imply that the supposed
robust events during which the eccentricities were excited must have
occurred at least $\sim10^5Q'/0.01yr$ ago, the time of the secular
evolution in this system.

\section{Conclusion and discussion}
\label{discuss}

We have investigated the possible evolution histories of the inner three planets in HD40307
system. We use the N-body model, adding tides raised by the star on
the planet and the general relativity as the additional forces. We
find three kinds of paths along which the system can evolve to the
current configuration. The three paths all need moderate
eccentricities ($\sim0.15$), which are supposed to result from some
robust events, such as collision or scattering. Moreover, the three
paths originate from different areas in the $P_2/P_1$ versus
$P_3/P_2$ phase space, while they all pass the apsidal alignment
duration before the final tidal equilibrium arrives.

Minimum masses are used in all cases above. We also made some cases
with twice minimum masses, and found that the stronger effects among
planets cause higher eccentricities. However, the actual evolution time is proportional
to the damping parameter $Q'$, and Neptune-like planets tend to have
bigger $Q'$ than Earth-like planets. As a result, the evolutions for
more massive planets should not be faster than the ones for minimum
masses. Another assumption we have made is that the three planets'
$Q'$ are the same or at least in the same magnitude, and it should be
most likely to be the truth due to their minimum masses in the same
magnitude. But in case this is not true, which means these planets
might have totally different components, then the evolution process
would be different from what we have discussed. All these are
waiting for a further detection.

\acknowledgments
This work has been supported by the National Basic Research Program of China (No. 2013CB834900), the National Natural Science Foundation of China under grant Nos. 11333002, 10925313, 10933004, the Strategic Priority Research Program "The Emergence of Cosmological Structures" of the Chinese Academy of Sciences, Grant No. XDB09000000, and the Minor Planet Foundation of Purple Mountain Observatory.

\clearpage







\end{document}